\documentclass[12pt]{article}

\usepackage[english]{babel}
\usepackage[utf8x]{inputenc}
\usepackage[T1]{fontenc}

\usepackage[margin=1in]{geometry}
\usepackage{graphicx}
\usepackage{setspace}
\usepackage{amssymb}
\usepackage{amsmath}
\usepackage{epstopdf}
\usepackage[numbers, sort&compress]{natbib}
\usepackage{authblk}

\usepackage{xcolor}

\usepackage{standalone}

\newcounter{Box}

\newlength{\standardskip}
\makeatletter
\newcommand{\@minipagerestore}{\setlength{\parskip}{\standardskip}}
\makeatother

\title{Linking evolutionary and ecological theory illuminates
  non-equilibrium biodiversity \vspace{2em}}

\author[1, 2]{A. J. Rominger}
\author[3]{I. Overcast}
\author[1]{H. Krehenwinkel}
\author[1]{R. G. Gillespie}
\author[1, 4]{J. Harte}
\author[3]{M. J. Hickerson}

\affil[1]{Department of Environmental Science, Policy and Management,
  University of California, Berkeley}
\affil[2]{Santa Fe Institute}
\affil[3]{Biology Department, City College of New York}
\affil[4]{Energy and Resource Group, University of California,
  Berkeley}

\date{}

\begin{document}
\maketitle
\thispagestyle{empty}
\addtocounter{page}{-1}

\noindent
{\it Corresponding author:} Rominger, A.J. (ajrominger@gmail.com).


\pagebreak

\section*{Abstract}

Whether or not biodiversity dynamics tend toward stable equilibria
remains an unsolved question in ecology and evolution with important
implications for our understanding of diversity and its
conservation. Phylo/population genetic models and macroecological
theory represent two primary lenses through which we view
biodiversity. While phylo/population genetics provide an averaged view
of changes in demography and diversity over timescales of generations
to geological epochs, macroecology provides an ahistorical description
of commonness and rarity across contemporary co-occurring species. Our
goal is to combine these two approaches to gain novel insights into
the non-equilibrium nature of biodiversity.  We help guide near future
research with a call for bioinformatic advances and an outline of
quantitative predictions made possible by our approach.

\pagebreak

\section{Non-equilibrium, inference, and theory in ecology and evolution}

The idea of an ecological and evolutionary equilibrium has pervaded
studies of biodiversity from geological to ecological, and from global
to local \citep{sepkoski1984, rabosky2009, hubbell2001, harte2011,
  chesson2000, tilman2004}. The consequences of non-equilibrium
dynamics for biodiversity are not well understood and the need to
understand them is critical with when anthropogenic pressures forcing
biodiversity into states of rapid transition
\citep{blonder2015}. Non-equilibrial processes could profoundly inform
conservation in ways only just beginning to be explored
\citep{wallington2005}.
  
Biodiversity theories based on assumptions of equilibrium, both
mechanistic \citep{hubbell2001, chesson2000, tilman2004} and
statistical \citep[see the Glossary;][]{harte2011, pueyo2007}
have found success in predicting ahistorical patterns of diversity
such as the species abundance distribution \citep{white2012,
  hubbell2001, harte2011} and the species area relationship
\citep{hubbell2001, harte2011}. These theories assume a
macroscopic equilibrium in terms of these coarse-grained metrics, as
opposed to focusing on details of species identity \citep[such as
in][]{blonder2015}, although macroscopic and microscopic approaches
are not mutually exclusive.  Nonetheless, the equilibrium assumed by
these theories is not realistic \citep{ricklefs2006}, and many
processes, equilibrial or otherwise, can generate the same
macroscopic, ahistorical predictions \citep{mcgill2007}.

Tests of equilibrial ecological theory alone will not allow us to
identify systems out of equilibrium, nor permit us to pinpoint the
mechanistic causes of any non-equilibrial processes. The dynamic
natures of evolutionary innovation and landscape change suggest that
ecological theory could be greatly enriched by synthesizing its
insights with inference from population genetic theory that explicitly
accounts for history. This would remedy two shortfalls of equilibrial
theory: 1) if theory fits observed ahistorical patterns but the
implicit dynamical assumptions were wrong, we would make the wrong
conclusion about the equilibrium of the system; 2) if theories do not
fit the data we cannot know why unless we have a perspective on the
temporal dynamics underlying those data.

No efforts to date have tackled these challenges. We propose that
combining insights from ecological theory and inference of
evolutionary and demographic change from genetic data will allow us to
understand and predict the consequences of non-equilibrial processes
in governing the current and future states of ecological
assemblages. The advent of next generation sequencing (NGS) approaches to
biodiversity, from microbes to arthropods \citep{taberlet2012,
  gibson2014, shokralla2015, ji2013, zhou2013, bohmann2014,
  linard2015, leray2015, dodsworth2015, liu2016, venkataraman2015}
have made unprecedented data available for synthesizing insights form
ecological theory and genetics/genomics.  However, we need a tool set
of bioinformatic methods (Box \ref{box:dry}) and meaningful
predictions (section \ref{sec:pred}) grounded in theory to make use of
those data.
%
% Data will take the form of both standard ecological
% metrics such as species abundances, as well as summaries of
% demographic/diversity dynamics inferred from genetics. Theory-based
% predictions will consist of connecting deviations from ecological
% theory and regions of parameter space with the dynamic processes
% inferred from genetics, all aided by new bioinformatic advances. 
%
We present the foundation of this tool set here.

\section{Ecological theories and non-equilibrium}

Neutral and statistical theories in ecology focus on macroscopic
patterns, and equilibrium is presumed to be relevant to those
patterns, but not the finer-grained properties of
ecosystems. Our goal throughout is not to validate neutral or statistical
theories---quite the opposite, we propose new data dimensions, namely
genetics, to help better test alternative hypotheses against these
null theories, thereby gaining insight into what non-neutral and
non-statistical mechanisms are at play in systems of interest.

Non-neutral and non-statistical models \citep[e.g.,][]{tilman2004,
  chesson2000} also invoke ideas of equilibrium in their
derivation. However, these equilibria focus on the micro-scale details
of species interactions and therefore do not fall within our primary
focus, and could in fact be drivers of non-equilibrium and thus
interesting alternative hypotheses to test.  We focus explicitly on
simple yet predictive theories for their utility as null models, not
because of a presumption of their realism.

To use these theories as null models, we need a robust measure of
goodness of fit. The emerging consensus is that likelihood-based test
statistics should be preferred \citep{baldridge2016}. The ``exact
test'' of Etienne \citep{etienne2007} has been extended by Rominger
and Merow \citep{meteR} into a simple z-score which can parsimoniously
describe the goodness of fit between theory and pattern.  We advocate
its use in our proposed framework.

The neutral theory of biodiversity \citep[NTB;][]{hubbell2001} is a
useful null because it assumes that one mechanism---demographic
drift---drives community assembly.  Equilibrium occurs when
homogeneous stochastic processes of birth, death, speciation and
immigration have reached stationarity. Thus neutrality in ecology is
analogous to neutral drift in population genetics \citep{hubbell2001}.

Rather than assuming any one mechanism dominates the assembly of
populations into a community, statistical theories assume all
mechanisms could be valid, but their unique influence has been lost to
the enormity of the system and thus the outcome of assembly is a
community in statistical equilibrium \citep{harte2011, pueyo2007}. The
mechanistic agnosticism is what makes statistical theories useful
nulls. These statistical theories are also consistent with niche-based
equilibria \citep{pueyo2007, neill2009} if complicated, individual or
population level models with many mechanistic drivers were to be
upscaled to entire communities. 

The maximum entropy theory of ecology \citep[METE;][]{harte2011}
derives its predictions by condensing the many bits of mechanistic
information down into ecological state variables and then
mathematically maximizing information entropy conditional on those
state variables. METE can predict multiple ahistorical patterns,
including distributions of species abundance, body size, spatial
aggregation, and trophic links \citep{harte2011, rominger2015}, making
for a stronger null theory \citep{mcgill2003}. However, multiple
dynamics can still map to this handful of metrics \citep{mcgill2007}
and while extensive testing often supports METE's predictions
\citep{harte2011, white2012, xiao2015} at single snapshots in time,
METE fails to match observed patterns in disturbed and rapidly
evolving communities \citep{rominger2015, harte2011}. We cannot know
the cause of these failures within the current framework of
equilibrium theory testing without adding metrics that capture
temporal dynamics.

\section{Inferring non-equilibrium dynamics}

Unlocking insight into the dynamics underlying community assembly is
key to overcoming the limitations of analyzing ahistorical patterns
with equilibrial theory. While the fossil record could be used for
this task, it has limited temporal, spatial, and taxonomic
resolution. Here we instead focus on population/phylogenetic insights
into rates of change of populations and species because of the
detailed characterization of demographic fluctuations, immigration,
selection, and speciation they provide. Bridging ecological theory
with models from population/phylogenetics has great potential
\citep{webb2002, lavergne2010, mcgaughran2015, laroche2015,
  papadopoulou2011, dexter2012} that has yet to be fully realized. How
we can best link the inferences of change through time from
population/phylogenetics with inferences from macroecology is governed
by what insights we can gain from genetic perspectives on demography
and diversification.

Coalescent theory \citep{kingman1982stochasti, rosenberg2002} is one
of the fundamental population genetics tools allowing model-based
estimation of complex historical processes. These include population
size fluctuations \citep{kuhner1998}, divergence and/or colonization
times \citep{charlesworth2010, edwards2000}, migration rates
\citep{wakeley2008}, selection \citep{kern2016}, and complex patterns
of historical population structure \citep{prado-martinez2013} and gene
flow \citep{beerli2001, hey2004}. This approach can also be put in a
multi-species, community context via hierarchical demographic models
\citep{xue2015, hickerson2006, carstens2016, chan2014}, even when only
small numbers of genetic loci are sampled from populations
\citep{drummond2005}.

These modeled demographic deviations from neutral demographic
equilibrium can also be condensed into multi-species summary
statistics. For example, Tajima's D, which measures the strength of
non-equilibrium demography in a single population \citep[see Glossary
for more details;][]{tajima1989, jensen2005, schrider2016,
  stephan2016}, could be averaged over all populations in a sample.

\section{Current efforts to integrate evolution into ecological theory} \label{sec:toDate}

While quantitatively integrating theory from ecology, population
genetics, and phylogenetics has not yet been achieved, existing
efforts to synthesize perspectives from evolution and ecology point
toward promising directions despite being hindered by one or more
general issues: 1) lack of a solid theoretical foundation, 2)
inability to distinguish multiple competing alternative hypotheses, 3)
lack of comprehensive genetic/genomic data, and 4) lack of
bioinformatic approaches to resolve species and their abundances.

Phylogenetic information has been incorporated into studies of the NTB
to better understand its ultimate equilibrium \citep{jabot2009,
  burbrink2015}.  However, phylogenetic reasoning also points out the
flaws in the NTB's presumed equilibrium \citep{ricklefs2006}.
Attempts to correct the assumed dynamics of NTB through ``protracted
speciation'' \citep{rosindell2010} are promising, and while their
implications for diversification have been considered
\citep{etienne2011}, these predictions have not been integrated with
demographic and phylogeographic approaches
\citep[e.g.,][]{charlesworth2010, edwards2000, prado-martinez2013}
that have the potential to validate or falsify presumed mechanisms of
lineage divergence.  Such demographic studies, particularly
phylogeographic investigations of past climate change
\citep{smith2012, hickerson2005}, have highlighted the non-equilibrium
responses of specific groups to perturbations that must be confronted
by ecological theory, but no attempt has been made to scale up these
observations to implications at the level of entire communities. The
recent growth in joint studies of genetic and species diversity
\citep{vellend2005amnat, papadopoulou2011} have been useful in linking
population genetics with ecological and biogeographic concepts. These
correlative studies could be bolstered by developing full joint models
that link community assembly, historical demography and
coalescent-based population genetics combined with NGS.

Studies have also used chronosequences or the fossil record in
combination with neutral and/or statistical theory to investigate
changes over geologic time in community assembly mechanisms
\citep{wagner2006, rominger2015}. While these
studies have documented interesting shifts in assembly mechanisms,
including departures from equilibrium likely resulting from
evolutionary innovations, understanding exactly how the evolution of
innovation is responsible for these departures cannot be achieved
without more concerted integration with genetic data.

\section{What is needed now}

A key limitation to using ahistorical theory to infer dynamic
mechanisms is that multiple mechanisms, from simple and equilibrial to
complex, can map onto the same ahistorical pattern
\citep{engen1996lnorm, mcgill2003}. This means that even when a theory
describes the data well, we do not really know the dynamics that led
to that good fit \citep{ricklefs2006}.

Quantitatively integrating the dynamics inferred from population and
phylogenetic approaches with ahistorical, equilibrial ecological
theory can break this many-to-one mapping of mechanism onto prediction
and contextualize whether a match between ahistorical pattern and
theory truly results form equilibrial dynamics or only falsely
appears to. There are two complementary approaches to achieve this
integration (both discussed further in Box \ref{box:dry}):

\begin{itemize}
\item Option 1: using dynamics from genetic inference to predict and
  understand deviations from ahistorical theories. This amounts to
  separately fitting ahistorical theory to typical macroecological
  data, while also fitting population genetic and/or phylogenetic
  models to genetic data captured for the entire community. Doing so
  requires substantial bioinformatic advances that would allow the
  joint capture of genetic or genomic data from entire community
  samples using NGS, while also estimating accurate abundances from
  NSG output. Separating model fitting avoids assumptions about how
  macroecological quantities like abundance scale to evolutionary
  metrics like effective population size; however, this approach does
  not facilitate the elegance of model comparison as does joint modeling.
\item Option 2: building off existing theories, develop new joint
  models that simultaneously predict macroecological and population
  genetic patterns. This amounts to building hierarchical models that
  take genetic data as input and integrate over all possible community
  states that could lead to these genetic data given a model of
  community assembly and a model of population coalescence. This
  approach requires making assumptions about how abundances scale to
  effective population sizes, but is better suited for comparing
  competing joint models of evolutionary history and assembly.
\end{itemize}

\subsection{What we gain from this framework}

Using our proposed framework, we can finally understand why
ahistorical theories fail when they do---is it because of rapid
population change, or evolution/long-distance dispersal of novel
ecological strategies? We could predict whether a system that obeys
the ahistorical predictions of equilibrial ecological theory is in
fact undergoing major non-equilibrial evolution. We could better
understand and forecast how/if systems out of equilibrium are likely
to relax back to equilibrial patterns. With such a framework we could
even flip the direction of causal inference and understand ecological
drivers of diversification dynamics. This last point bears directly on
long-standing debates about the importance of competitive limits on
diversification\citep{rabosky2009, harmon2015amNat}. Conclusions about
phylogenetic patterns (e.g. diversification slowdowns) would be more
believable and robust if combined with population genetic inference
(e.g. declining populations) and community patterns (e.g.  deviation
from equilibrium).

\section{Evo-ecological predictions for systems out of equilibrium} \label{sec:pred}

\begin{figure}[!hbp]
  \centering
  \includegraphics[scale=1]{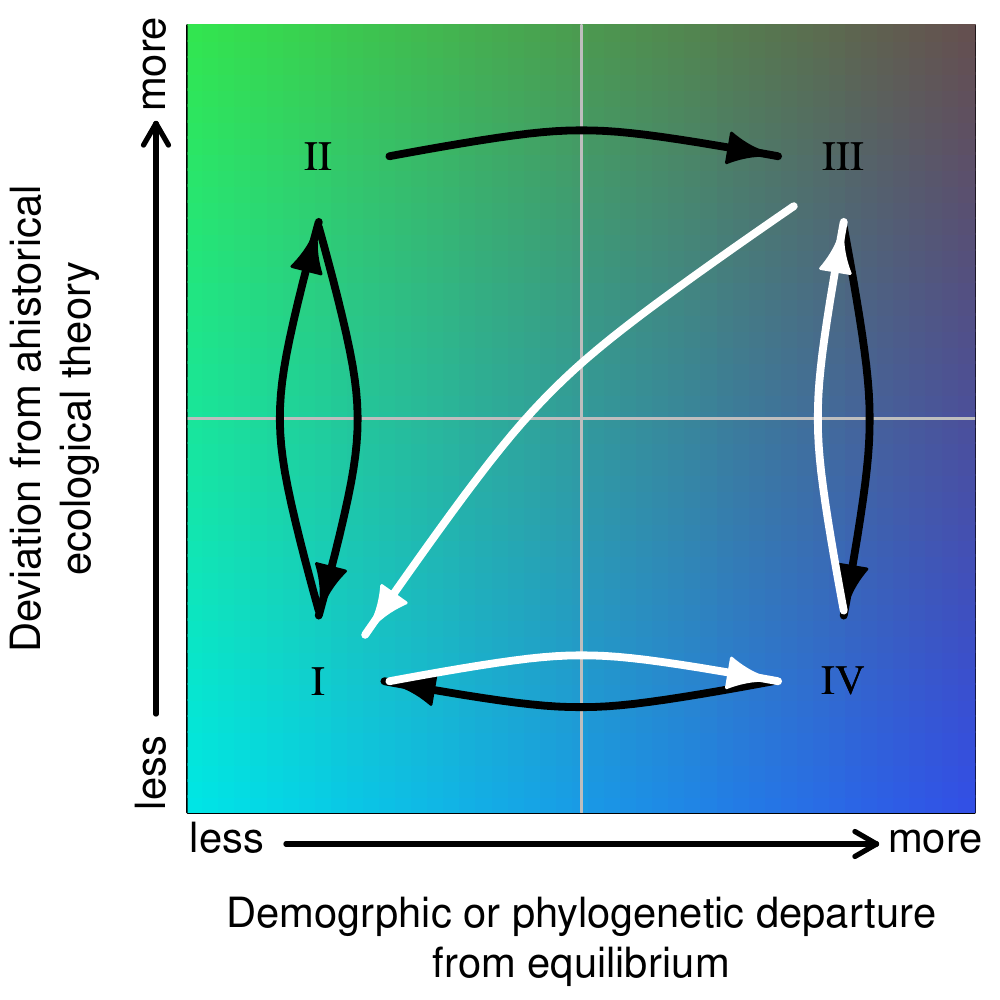}
  \caption{Hypothesized cycles between different states of equilibrium
    and non-equilibrium in ecological theory (y-axis) and evolutionary
    demography/diversification (x-axis). Deviations from ecological
    theory can be quantified by the previously discussed exact tests
    \citep{etienne2007} and z-scores \citep{meteR}, while many
    statistics are available to quantify departure from
    demographic/diversification steady state including the previously
    discussed Tajima's D. Panels I--IV are discussed in the text.
    Colors correspond to deviation from ahistorical ecological theory
    and evolutionary equilibrium.  Black cycle corresponds to
    non-equilibrium initiated by ecological disturbance (with
    potential to continue to evolutionary non-equilibrium or
    relaxation to equilibrium). White cycle is initiated by
    evolutionary innovation.}
  \label{fig:cycles}
\end{figure}

\begin{figure}[!hbp]
  \centering
  \includegraphics[scale=1]{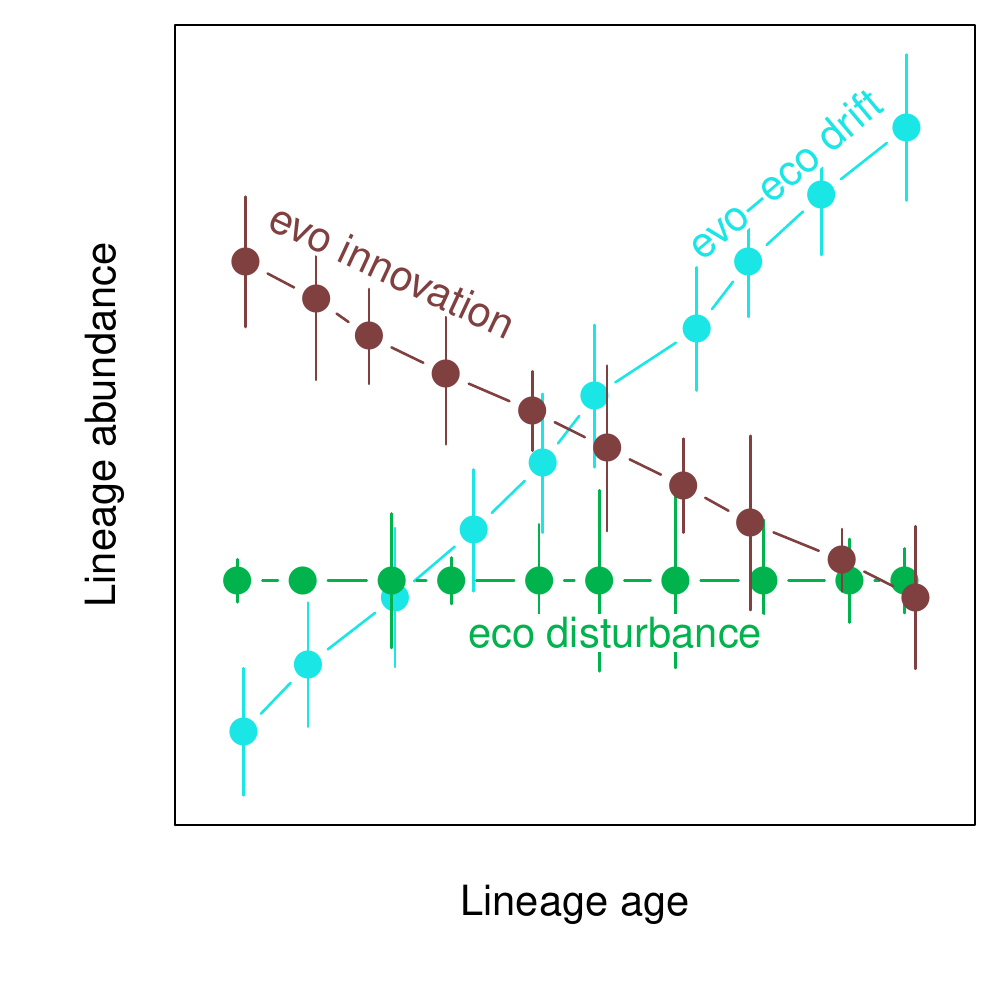}
  \caption{Hypothesized relationships between lineage age and
    abundance under different evo-ecological scenarios. Colors
    correspond to panels in Figure \ref{fig:cycles}: teal is
    evo-ecological equilibrium; green is rapid transition to
    ecological non-equilibrium following short timescale disturbance;
    dark brown is non-equilibrium in both ecological and evolutionary
    metrics.}
  \label{fig:age-abund}
\end{figure}

We propose a simple yet powerful way to summarize joint inference into
deviations from ecological and evolutionary/demographic
equilibrium. Assuming that equilibrial models have been independently
or jointly fit (Box \ref{box:dry}) to a dataset of macroecological
metrics (such as species abundances) and genetic/genomic variables
(such as community-wide polymorphism data) we can then contrast the
system's deviation from ecological equilibrium with its deviation from
evolutionary/demographic equilibrium
(Fig. \ref{fig:cycles}). Ecological deviations can be measured by,
e.g., the previously discussed z-score \citep{meteR}, while
evolutionary/demographic deviations can be captured by summary
statistics such as community-averaged Tajima's D. The space of
equilibrium and non-equilibrium states that this comparison generates
(Fig. \ref{fig:cycles}) can be used to understand a communitie's
current state, and predict its past and future. Additional predictions
from joint eco-evolutionary inference can be tested to further
understand a systems' trajectory through phases of equilibrium and
non-equilibrium. One particularly useful metric is the relationship
between lineage age (colonization or divergence time inferred from
molecular data) and lineage abundance (Fig. \ref{fig:age-abund}),
which is known to be a telling test of the NTB \citep{ricklefs2006,
  rosindell2010}.  Cometing different models of assembly and
coalescence in a model selection framework can also provide insight.

\subsection{Cycles of non-equilibrium}

Ecosystems experience regular disturbances which can occur on
ecological time-scales, such as primary succession, or evolutionary
time scales, such as evolution of novel innovations that lead to new
ecosystem processes \citep{erwin2008}. We hypothesize that these
regular disturbances can lead to cycles of non-equilibrium in observed
biodiversity patterns. Using the phase space of equilibrium and
non-equilibrium states showing in Figure \ref{fig:cycles} a clockwise
cycle through this space would indicate:

\begin{itemize}
\item Panel I $\rightarrow$ II: following rapid ecological
  disturbance, ecological metrics diverge from equilibrium. The system
  could potentially relax back to equilibrium (Panel II $\rightarrow$
  Panel I), indicating a stationary disturbance process that has no
  net evolutionary consequences. Conversely,
\item Panel II $\rightarrow$ III: ecological non-equilibrium spurs
  evolutionary non-equilibrium leading to both ecological and
  evolutionary metrics diverging from equilibrium values
\item Panel III $\rightarrow$ IV: evolutionary innovations provide the
  means for ecological processes to
  re-equilibrate to their environments
\item Panel IV $\rightarrow$ I: finally a potential return to
  equilibrium on both ecological and evolutionary time scales once
  evolutionary processes have also relaxed.
\end{itemize}

Cycles through this space could also occur in a counterclockwise
direction, being initiated by an evolutionary innovation. Under such a
scenario we hypothesize the cycle to proceed:

\begin{itemize}
\item Panel I $\rightarrow$ IV: non-equilibrium evolution (including
  sweepstakes dispersal) leading to departure from evolutionary
  equilibrium before departure from ecological equilibrium
\item Panel IV $\rightarrow$ III: non-equilibrial ecological response
  to non-equilibrium evolutionary innovation
\item Panel III $\rightarrow$ I: ecological and evolutionary
  relaxation
\end{itemize}

We hypothesize that the final transition will be directly to a joint
equilibrium in ecological and evolutionary metrics (Panel I) because a
transition from panel III to panel II is unlikely, given that
ecological rate are faster than evolutionary rates. In general these
cycles can be combined arbitrarily depending on the forces and
dynamics present in the system.  However, transitions where
evolutionary rates must opperate faster than ecological rates
(i.e. Panel III $\rightarrow$ II and Panel IV $\rightarrow$
Panel II) are less likely.

A complete cycle cannot be observed without a time machine, but by
combining ahistorical ecological theory and population/phylogenetic
inference methods using community-level genetic data we can identify
where in this space our focal systems are located. To determine their
trajectory through this space we must more deeply explore the joint
inference of community assembly and evolutionary processes. In the
following sections we do that for each transition shown in Figure
\ref{fig:cycles} using patterns of lineage age and abundance together
with model selection.

\subsection{Systems undergoing rapid ecological change}

For systems whose metrics conform to demographic equilibrium, but
deviate from equilibrial ecological theory (Panel I $\rightarrow$ II),
a lack of correlation between lineage age and lineage abundance would
indicate that rapid ecological change underlies their dynamics. If
slow, equilibrial evolutionary drift is punctuated by regular
ecological perturbations, population size would be independent of age
(Fig. \ref{fig:age-abund}). Actual abundance should similarly be
uncorrelated with effective population size in joint genetic-assembly
models.

\subsection{Ecological relaxation}

Ecological relaxation occurs when populations return to steady
state. Both ecological (Panel II $\rightarrow$ I) and evolutionary
(Panel III $\rightarrow$ IV) mechanisms can facilitate this process
(e.g. changes in local population sizes following environmental change
\citep{blonder2015}, or evolution of new species interactions such as
host switching \citep{rominger2015}). If ecological mechanisms are
responsible, age and abundance should again be uncorrelated; if
evolutionary mechanics are responsible, age and abundance should be
negatively correlated (Fig. \ref{fig:age-abund}).

\subsection{Non-equilibrium ecological communities fostering non-equilibrium evolution}

A lack of equilibrium in an ecological assemblage means that the
system will likely experience change in order to re-equilibrate. If
ecological relaxation does not occur---by chance, or because no
population present is equipped with the adaptations to accommodate the
new environment---then the system is open to evolutionary innovation
(Panel II $\rightarrow$ III).  Such innovation could take the form of
elevated speciation or long-distance immigration. Speciation and
sweepstakes immigration/invasion will yield very different
phylogenetic signals, however, their population genetic signals in a
non-equilibrium community may be very similar (e.g. rapid population
expansion). Thus, where non-equilibrium communities foster
non-equilibrium diversification (either through speciation or
invasion) we expect to see a negative relationship between lineage age
and abundance (Fig.  \ref{fig:age-abund}). Similarly, in a joint
genetic-assembly modeling framework, population expansion models
should be favored over demographically stationary models.

\subsection{Non-equilibrium evolution fostering non-equilibrium ecological dynamics}

If evolutionary processes, or their counterpart in the form of
sweepstakes immigration/invasion, generate new ecological strategies
in a community, this itself constitutes a form of disturbance pushing
the system to reorganize (Panel IV $\rightarrow$ III).  Evolutionary
change would have to be extremely rapid to force ecological metrics
out of equilibrium, thus in this scenario we would expect a negative
correlation between age and abundance (Fig. \ref{fig:age-abund}) and
model selection favoring joint genetic-assembly models with highly
structured populations and rapid divergence rates. If data become
available for large regions of genomes for entire communities, signals
of strong selection could also validate non-equilibrium evolution
fostering non-equilibrium ecological dynamics (see {\bf Outstanding
  Questions}).

\subsection{Evolutionary relaxation}

Evolutionary demographic models average over timescales determined by
generation time, population size and mutation/selection balance
\citep{kingman1982stochasti, rosenberg2002}. Evolutionary relaxation
(Panel III $\rightarrow$ I or IV $\rightarrow$ I) means this
time-averaged history returns to stationarity, which can occur if
perturbations are absent, or occur on rapid enough time scales
(i.e. Panel II $\rightarrow$ I) to be averaged over. By definition, if
a system is found in evolutionary demographic equilibrium it has
forgotten any non-equilibrium phases in its history.  Thus to detect
this kind of long-term relaxation we need data from the fossil record
(see {\bf Outstanding Questions}).

\section{Harnessing evo-ecological measures of non-equilibrium for a changing world}

Inference of community dynamics needs to expand the dimensions of data
it uses to draw conclusions about assembly of biodiversity and
whether/when this is an equilibrial or non-equilibrial process. The
need for more data dimensions is supported by others
\citep{mcgill2007}; however, accounting for the complexities of
history by explicitly linking theories of community assembly with
theories of evolutionary genetics is novel, and made possible by
advances in:
\begin{enumerate}
\item high throughput sequencing (Box \ref{box:wet}) that allow
  genetic samples to be economically and time-effectively produced on
  unprecedented scales
\item bioinformatic methods (Box \ref{box:dry}) that allow
  us to make sense of these massive community-wide genetic/genomic
  datasets
\item theory development (section \ref{sec:pred}) that provides
  meaningful predictions to test with our new bioinformatic approaches
\end{enumerate}

This framework is a fertile cross pollination of two fields that,
while successful in their own right, are enhanced by their
integration. While comparative historical demographic models are
advancing \citep{xue2015, hickerson2006, carstens2016, chan2014},
testing community-scale hypotheses with multi-taxa data would be
profoundly improved and enriched if population genetic model were
grounded in macroecological theory.  What is more, models of community
assembly have been overly reliant on ahistorical patterns and
assumptions of equilibrium, which are by themselves often insufficient
for distinguishing competing models of assembly \citep{mcgill2007}.
The field is ready to fully merge these two approaches using the wet
lab, bioinformatic, and theoretical approaches we advocate here . The
time to do so is now, as society faces an increasingly non-equilibrium
world, challenging our fundamental understanding of what forces
govern the diversity of life and how we can best harmonize human
activities with it.

\section*{Acknowledgements}

We would like to thank L. Schneider for helpful comments. AJR
acknowledges funding from the Berkeley Initiative in Global Change
Biology and NSF DEB-1241253.

\pagebreak

\bibliographystyle{tree}
\bibliography{references.bib}

\pagebreak

\section*{Boxes}

\refstepcounter{Box}\label{box:wet}
\subsection*{Box \theBox: Wetlab techniques}
    
Next generation sequencing (NGS) technology has ushered in a revolution in
evolutionary biology and ecology. The large scale recovery from bulk
samples (e.g. passive arthropod traps) of species richness, food web
structure, and cryptic species promise unprecedented new insights into
ecosystem function and assembly \citep{krehenwinkel2016,
  shokralla2015, gibson2014, taberlet2012}.  Two approaches, differing
in cost and effectiveness, have emerged.

\paragraph{Metabarcoding} describes the targeted PCR amplification and
next generation sequencing of short DNA barcode markers (typically
~300-500 bp) from community samples \citep{ji2013}. The resulting
amplicon sequences can be clustered into OTUs or grafted onto more
well supported phylogenies. Even minute traces of taxa in
environmental samples can be detected using metabarcoding
\citep{bohmann2014}.  Amplicon sequencing is cheap, requires a small
workload, and thus allows rapid inventories of species composition and
species interactions in whole ecosystems \citep{gibson2014,
  leray2015}. However, the preferential amplification of some taxa
during PCR leads to highly skewed abundance estimates
\citep{elbrecht2015} from metabarcoding libraries.

\paragraph{Metagenomic approaches}, in contrast, avoid marker specific
amplification bias by sequencing libraries constructed either from
untreated genomic DNA \citep{dodsworth2015, linard2015}, or
after targeted enrichment of genomic regions \citep{liu2016}. While
being more laborious, expensive and computationally demanding than
metabarcoding, metagenomics thus offers improved accuracy in detecting
species composition \citep{zhou2013}. Moreover, the
assembly of high coverage metagenomic datasets recovers large
contiguous sequence stretches, even from rare members in a community,
offering high phylogenetic resolution at the whole community level
\citep{coissac2016}. Due to large genome sizes and high genomic
complexity, metazoan metagenomics is currently limited to the
assembly of short high copy regions. Particularly mitochondrial
and chloroplast genomes as well as nuclear ribosomal clusters are
popular targets \citep{dodsworth2015, coissac2016}. In contrast,
microbial metagenomic studies routinely assemble complete genomes
and characterize gene content and metabolic pathways even from complex
communities \citep{nielsen2014}. This allows unprecedented insights
into functional genetic process underlying community assembly and
evolutionary change of communities to environmental stress.  Such
whole genome based community analysis is not yet feasible for
macroorganisms. However, considering the ever increasing throughput
and read length of NGS technology, as well as
growing number of whole genomes, it might well become a possibility in
the near future, opening up unprecedented new research avenues for
community ecology and evolution.

\refstepcounter{Box}\label{box:dry}
\subsection*{Box \theBox: Bioinformatic advances}

While species richness can be routinely identified by sequencing bulk
samples using high throughput methods, estimating species abundance
remains challenging \citep{elbrecht2015} and severely limits the
application of high throughput sequencing methods to many
community-level studies. We propose two complementary approaches to
estimate species abundance from high throughput data.  The first
approach estimates abundance free from any models of community
assembly, the second jointly estimates the parameters of a specific
assembly model of interest along with the parameters of a
coalescent-based population genetic model.

\paragraph{Model-free abundance estimation.} We propose a pipeline
(Fig. \ref{fig:abundPipeline}) where raw reads are generated and
assembled into a phylogeny using standard approaches, and potentially
aided by additionally available sequence data in a super tree or super
matrix approach. The numbers of sequences assigned to each terminal
tip are then used in a Bayesian hierarchical model which seeks to
estimate the true number of organisms representing each terminal tip,
accounting for sequencing biases originating from, e.g. primer
affinity and copy number differences between taxa.  Information on
phylogenetic relatedness can inform modeled correlations in biases
between taxa \citep[e.g. copy number is known to be phylogenetically
conserved at least in microbes]{angly2014}. This approach is
particularly tailored to metabarcoding data. In a potentially powerful
extension, and thanks to the proposed Bayesian framework, information
from sequencing experiments that seek to calibrate metabarcoding
studies \citep[e.g.,][]{krehenwinkel2016} can be used to
build meaningfully informative priors and improve model
accuracy. Through a simulation study (described in the supplement) we
show that true underlying abundances can be accurately estimated
(Fig. \ref{fig:abundEst}).

\paragraph{Joint inference of community assembly and population
  genetic models.} Coupling individual-based, forward-time models of
community assembly with backwards-time hierarchical multi-taxa
coalescent models permits inference about the values of the parameters
in both models. This framework is flexible enough to incorporate
multiple refugia, colonization routes, ongoing migration and both
neutral and deterministic processes of assembly on time scales of
hundreds of thousands of years (Fig. \ref{fig:gimmeSAD}). A
forthcoming implementation \citep[gimmeSAD$\pi$;][]{overcast} jointly
models a forward-time individual-based neutral community assembly
process \citep{rosindell2015} and corresponding expectations of
community level genetic diversity and divergence using the msPrime
coalescent simulator \citep{kelleher2016}. This has been accomplished
by rescaling the time dependent local abundance distributions into
time dependent effective population size distributions while allowing
for heterogeneity in migration and colonization rates. This simulation
model can be combined with random forest classifiers and hierarchical
ABC to enable testing alternative assembly models, including models
that have not yet reached their theoretical equilibria.

\subsection*{Glossary}

\paragraph{ahistorical} Patterns or theories which do not contain
information about the historical processes that gave rise to them

\paragraph{approximate Bayesian computation (ABC)}. A method of
calculating an approximate posterior sample of parameters in a complex model
whose likelihood function cannot be analytically solved by simulating
realizations of the model, computing summary statistics from those
realizations, and probabilistically accepting or rejecting the
parameter values leading to those summary statistics based on their
agreement with the observed statistics computed from the real data.

\paragraph{coalescent} A stochastic, backwards in time population
genetic model in which alleles in the sample are traced to their
ancestors under demographic models of interest.

\paragraph{equilibrium} Equilibrium is often reserved for systems in
thermodynamic equilibrium---which all life violates.  By
``biodiversity equilibrium'' we make an analogy to thermodynamics and
say that biodiversity is in equilibrium if its marcrosopic state
(e.g. richness of species abundance distribution, but not necessarily
specific species compositions) is steady, and across arbitrary
subsystems, the same steady state applies.

\paragraph{hierarchical model} A modeling approach that facilitates
complex hypotheses and causal relationships by allowing model
parameters at one level to be dependent on parameters at another
level.

\paragraph{statistical equilibrium} In the context of biodiversity, a
description of a steady state arrived at not by the force of one or a
few deterministic mechanisms but by the stationary, statistical
behavior of very large collections of mechanistic drivers acting on
large assemblages of organisms.

\paragraph{Tajima's D} A metric of non-stationary evolution computed
as the difference between two distinct derivations of theoretical
genetic diversity. If neutral molecular evolution in a constant
population holds, both derivations should be equal, and otherwise if
assumptions of constant demography and neutral selection are violated,
will be unequal.

\pagebreak

\section*{Figures}

\pagebreak

\section*{Box \ref{box:dry} figures}

\setcounter{figure}{0}
\renewcommand{\thefigure}{\Roman{figure}}

\begin{figure}[!hbp]
  \centering
  \includegraphics[scale=0.4]{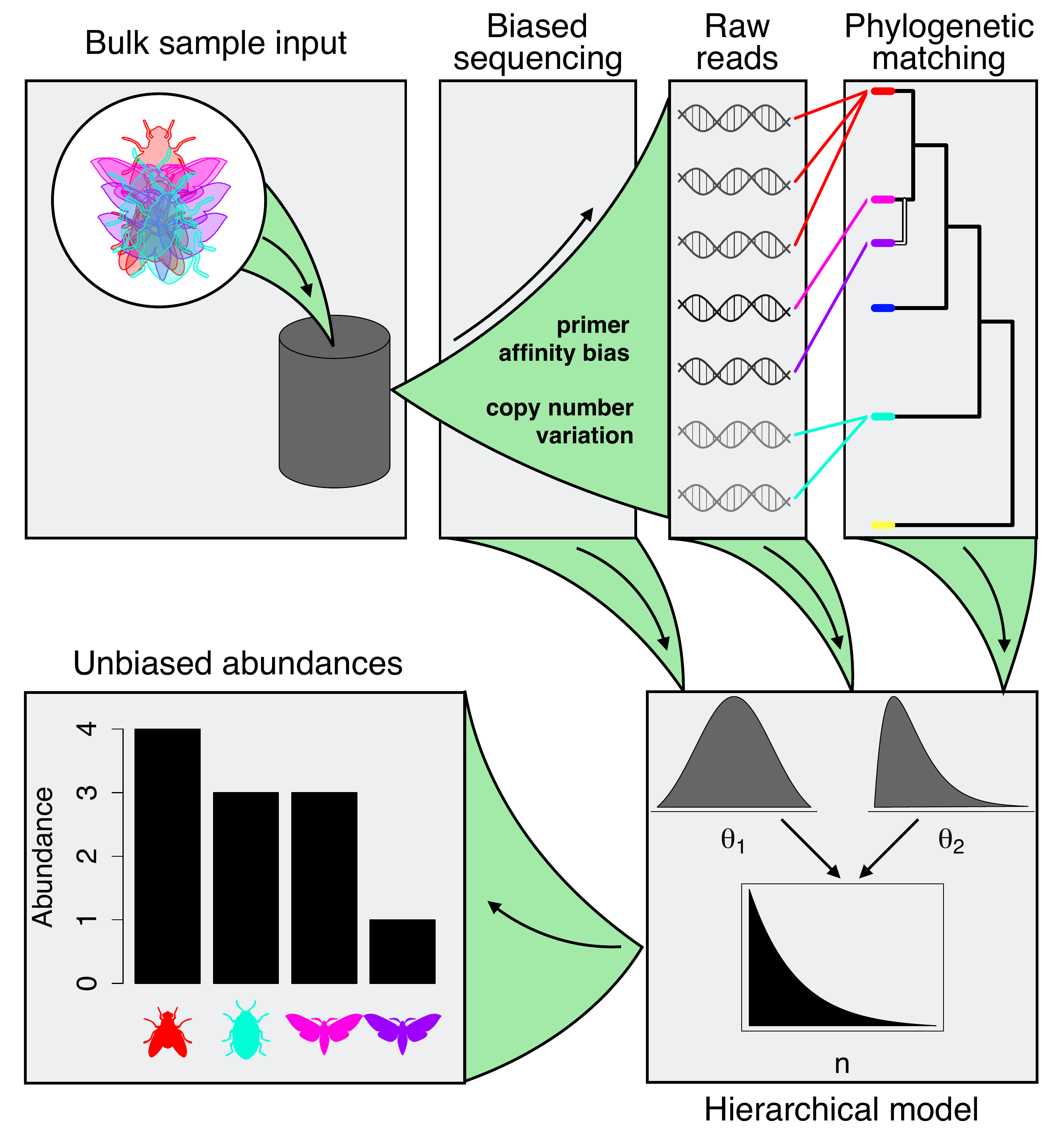}
  \caption{Pipeline to estimate true abundances from metabarcoding
    data. The pipeline follows sequence generation, matching sequences
    to a phylogeny (generated from the sequences themselves, or better
    yet from higher coverage data) and finally Bayesian hierarchical
    modeling leading to abundance estimates.}
  \label{fig:abundPipeline}
\end{figure}

\begin{figure}[!hbp]
  \centering
  \includegraphics[scale=1]{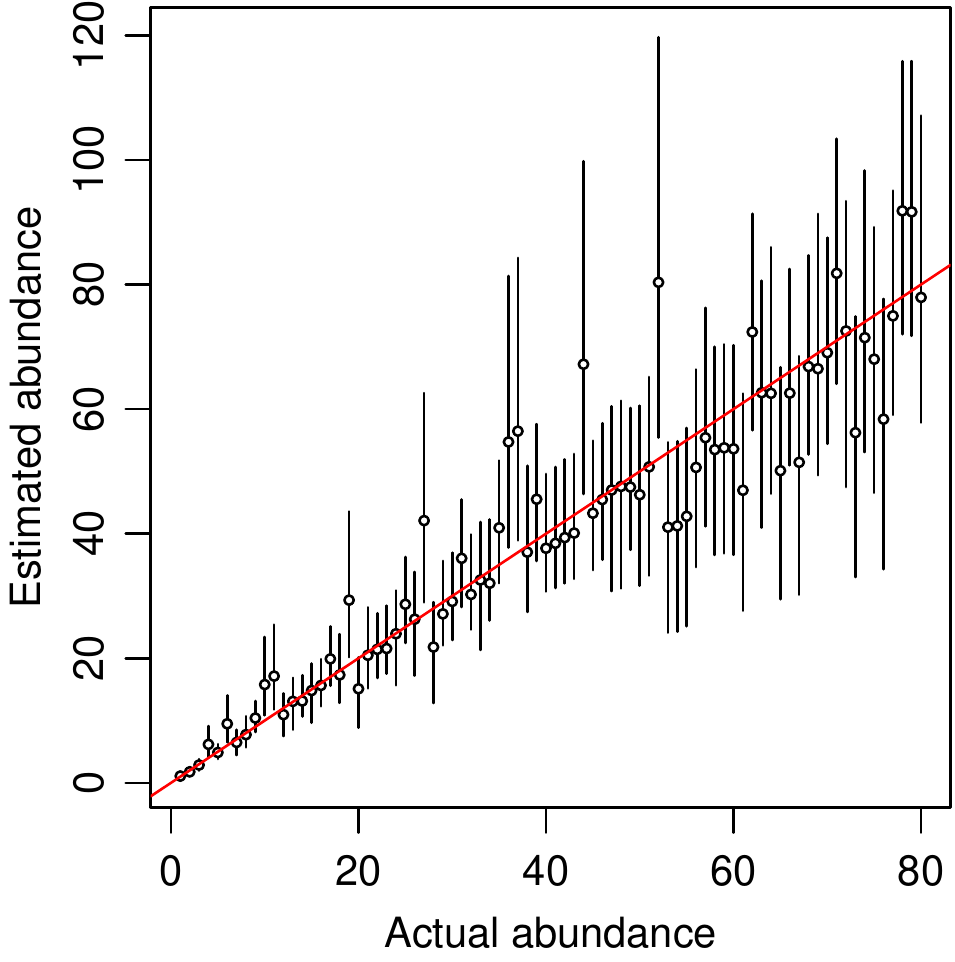}
  \caption{Demonstration of agreement between actual and estimated
    abundances. Actual (simulated) abundances are on the x-axis, which
    the y-axis shows estimated abundances (error bars are 95\% maximum
    credible intervals). The simulation study is described in the
    supplement.}
  \label{fig:abundEst}
\end{figure}

\begin{figure}[!hbp]
  \centering
  \includegraphics[scale=0.4]{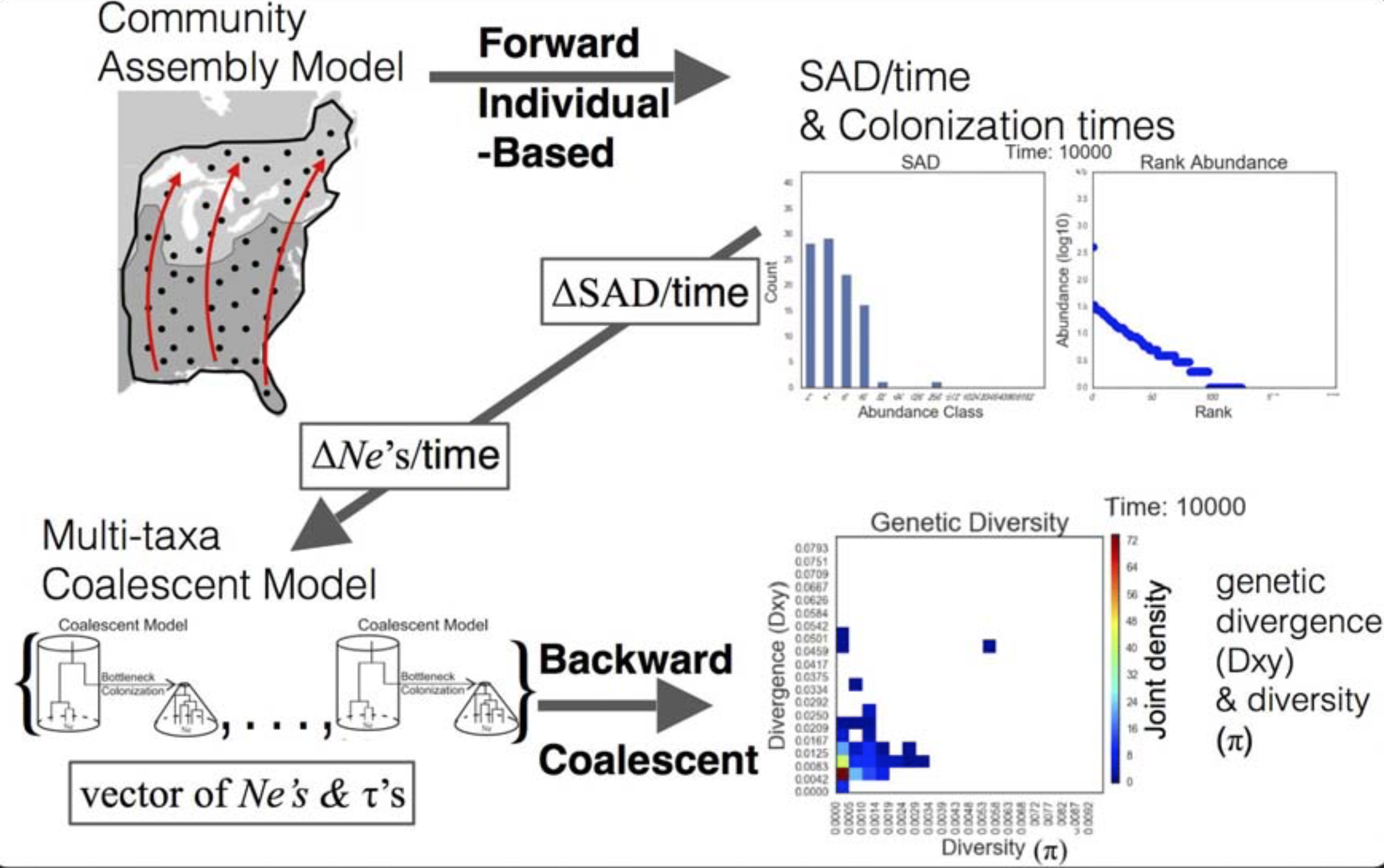}
  \caption{The gimmeSAD$\pi$ pipeline. The forward time models
    involves multi-regional expansion generating local abundance
    distributions over time with heterogeneity in
    colonization times. These temporally dynamic local abundances are
    re-scaled into local $n_e$ distributions over time to generate
    multi-species genetic data the the coalescent, which is summarized
    here with a time-dependent joint spectrum of genetic diversity
    statistics.}
  \label{fig:gimmeSAD}
\end{figure}

\end{document}